\documentclass[preprint,superscriptaddress,prb]{revtex4}
\usepackage{amssymb}

\usepackage{graphicx}
\usepackage{bm}
\usepackage[colorlinks=true,linkcolor=blue]{hyperref}

\expandafter\ifx\csname package@font\endcsname\relax\else
\expandafter\expandafter
\expandafter
\expandafter
\expandafter{\csname package@font\endcsname}
\fi
\begin{document}

\title{Unusual magnetoresistance in a topological insulator with a single
ferromagnetic barrier}

\author{B. D. Kong}
\affiliation{Department of Electrical and Computer Engineering, North Carolina State University, Raleigh, North Carolina 27695-7911}

\author{Y. G. Semenov}
\affiliation{Department of Electrical and Computer Engineering, North Carolina State University, Raleigh, North Carolina 27695-7911}

\author{C. Krowne}
\affiliation{U.S. Naval Research Laboratory, Washington, DC 20375}

\author{K. W. Kim}
\affiliation{Department of Electrical and Computer Engineering, North Carolina State University, Raleigh, North Carolina 27695-7911}

\begin{abstract}
Tunneling surface current through a thin ferromagnetic barrier on a
three-dimensional topological insulator is shown to possess an extraordinary
response to the orientation of barrier magnetization. In contrast to
conventional magnetoresistance devices that are sensitive to the relative
alignment of two magnetic layers, a drastic change in the transmission
current is achieved by a single layer when its magnetization rotates by 90
degrees. Numerical estimations predict a giant magnetoresistance as large as
800~\% at room temperature with the proximate exchange energy of 40~meV
at the barrier interface. When coupled with electrical control of magnetization direction, this phenomenon may be used to enhance the gating function with
potentially sharp turn-on/off for low power applications.
\end{abstract}

\pacs{73.20.-r,71.10.Pm,75.47.De,85.75.-d}
\maketitle

% insert suggested PACS numbers in braces on next line
% insert suggested keywords - APS authors don't need to do this
%\keywords{}

A topological insulator (TI) is a material that gapless fermion states on
its surface and inversion electronic bandgap in the bulk are conditioned by
exceptionally strong spin-orbital interaction.~\cite{Hasan2010,Zhang2009}
Recent experimental evidences of such surface states in three dimensional
materials~\cite{Hsieh2008} have sparked significant interests on the subject
from the point of view of both fundamental physics and potential benefits in
numerous device applications. In the immediate vicinity of the charge
neutrality point, the surface states represent an odd number of Dirac cones,
in which electrons can be described as massless Dirac fermions. Moreover,
the contact points of the conduction and valence bands are protected by the
time reversal symmetry that naturally implies suppression of back scattering
by nonmagnetic scatterers of arbitrary potential shape including the point
defects. This attribute fundamentally distinguishes the TIs from another
material with conic band structure, graphene.

For practical applications, an efficient way of controlling the surface
states is crucial. It is shown theoretically that TIs can have many exotic
physical properties related to the breaking of time reversal symmetry.~\cite%
{Hasan2010,Zhang2009} Of the possibilities, introduction of ferromagnetic materials is considered one particularly promising candidate. The exchange
interaction with a proximate magnetic film can affect the TI surface states
far more effectively than externally applied magnetic fields. The
state-of-the-art synthesis and fabrication technology available for magnetic
materials/structures also offers nanoscale controllability. Motivated by
this understanding, several studies have very recently examined their impact
on the electron transport properties in TIs.~\cite{Mondal2010,Yokoyama2010,Zhang2010,Burkov2010,Wu2010,Zhai2011} In most cases, multiple magnetic layers (or strips) were envisioned on a TI surface in a manner analogous to the conventional magnetoresistance devices. It is not surprising to learn that the modification of electron energy structures  induced by a combination of disparate magnetization directions can lead to a nontrivial response in electrical conduction. One drawback of this scenario, however, is that the functionality of the structure is rather limited to realize operations such as logic switching.

In this study, we exploit a different approach to the current control in a
TI with single conic surface states. Taking into account that the spin and
momentum variables are locked, electron transmission through a
\emph{single} ferromagnetic barrier can be significantly dependent on the
mutual orientation of the incident current $\mathbf{J}$ (or, the applied
electric field) and the magnetization $\mathbf{M}$ (see, for example, Fig.~1). For instance, in-plane magnetization of the ferromagnetic layer induces a shift in the Dirac cone of the surface states in contact away from the Brillouin zone center. When the displacement is in the direction of $\mathbf{J}$, the introduced magnetic barrier is not expected to affect electron transmission substantially as the conservation of transverse momentum can be maintained. In the general cases, however, the momentum conservation rule becomes restrictive, obstructing electron penetration in the barrier area. Moreover, orientating $\mathbf{M}$ perpendicular to the TI surface leads to a non-zero bandgap, i.e., an additional impediment to electron transmission. Thus, a \emph{single} ferromagnetic barrier with
variable magnetization appears to be sufficient to modulate the electrical
current on the surface of a TI.

The specific structure under consideration is schematically shown in Fig.~1. The TI surface is parallel to the $x$-$y$ plane and a thin strip of ferromagnetic insulator or dielectric (FMI) is deposited on top
along the $y$ axis. The FMI strip stretches from $x=0$ and $x=d$, a
width of $d$ that is shorter than the electron mean free path, while its y
dimension is assumed to be much larger. The current flow is designed to be in the x direction on the TI surface. The envisioned magnetoresistance effect can be analyzed, to sufficient accuracy, in terms of the surface state Hamiltonian
in the lowest order of momentum operator, which corresponds to
the conic dispersion relation. The quadratic~\cite{Culcer2010} and cubic
terms responsible for the hexagonal warping effects~\cite{Fu2009} are not
taken into account as the interest is primarily on the low energy electrons
near the Dirac point (i.e., small $E$). Accordingly the effective Hamiltonian is expressed as $H=\hbar v_{F}(\sigma _{x}p_{y}-\sigma_{y}p_{x})+V_g(x)+H_{ex}(x)$, where $\mathbf{p}$, $v_{F}$ and $\sigma _{i}$ are the in-plane wave vector of the electron, its Fermi velocity, and spin operators. $V_g(x)$ accounts for the electrostatic potential that may be applied through a gate electrode built on the FMI strip, causing $V_g(x)= V$ for $0\leq x\leq d$ and $V_g(x)=0$ elsewhere. In the mean-field approximation, the exchange interaction Hamiltonian with the proximate FMI takes the form $H_{ex}(x)=\alpha (x)\mathbf{M\cdot\sigma }$, where the factor $\alpha (x)$ is
proportional to the exchange integral in the surface region in contact ($0\leq x\leq d$) and zero otherwise. Combining $H_{ex}(x)$ with the first term results in simple momentum renormalization $k_{y}\rightarrow k_{y}+A^{ex}_{x}(x)$ and $k_{x}\rightarrow k_{x}-A^{ex}_{y}(x)$, where notations $\mathbf{k}=$ $\hbar v_{F}\mathbf{p}$ and $\mathbf{A}^{ex}=\alpha (x)\mathbf{M}$ are used. In this study, the magnetic field induced by the FMI strip is not taken into account since its impact (typically $\sim 0.1$~meV in units of energy) is negligible compared to the exchange interaction. Thus, it essentially resembles a one-dimensional potential barrier problem with the solutions representing the incident ($\psi _{I}$), barrier ($\psi _{B}$) and transmitted ($\psi _{T}$) regions.

In the case of in-plane directed $\mathbf{M}$, the electron transmission problem (with energy $E$) is solved in the form:
\begin{eqnarray}
\psi _{I}=\left(
\begin{array}{c}
\frac{ik_{x}+k_{y}}{E} \\
1\end{array}
\right) \frac{e^{ik_{x}x+ik_{y}y}}{\sqrt{2}}+r\left(
\begin{array}{c}
\frac{-ik_{x}+k_{y}}{E} \\
1\end{array}
\right) \frac{e^{-ik_{x}x+ik_{y}y}}{\sqrt{2}} \,, \nonumber
\end{eqnarray}
\begin{eqnarray}
\psi _{B}=a\left(
\begin{array}{c}
\frac{i(k^B_x-A^{ex}_{y})+(k_{y}+A^{ex}_{x})}{E-V} \\
1\end{array}\right) \frac{e^{{ik^B_x}x+ik_{y}y}}{\sqrt{2}}+b\left(
\begin{array}{c}
\frac{-i(k^{B}_{x}-A^{ex}_{y})+(k_{y}+A^{ex}_{x})}{E-V} \\
1\end{array}\right) \frac{e^{-i{k^B_x}x+ik_{y}y}}{\sqrt{2}}\,, \label{eqn:psi}
\end{eqnarray}
\begin{eqnarray}
\psi _{T}=t\left(
\begin{array}{c}
\frac{ik_{x}+k_{y}}{E} \\
1\end{array}\right) \frac{e^{ik_{x}x+ik_{y}y}}{\sqrt{2}}\,, \nonumber
\end{eqnarray}
where conservation of energy imposes $E^{2}=k_{x}^{2}+k_{y}^{2}$ for $\psi
_{I}$ and $\psi _{T}$, and ($E-V)^{2}=(k^B_x-A^{ex}_{y})^{2}+(k_{y}+A^{ex}_{x})^{2}$
for $\psi _{B}$. Matching the $\psi $ functions at $x=0$ and $x=d$ plays the
role of boundary conditions, defining coefficients $t$, $r$, $a$, and $b$.
Then, the transmission probability $\mathrm{P_{T}}=|t|^{2}$ is found for an
electron as a function of energy $E$ and incident angle. A similar procedure
can be applied to the case of $M_{z}\neq 0$. Under the ballistic transport
assumption, the conductance (per unit length in the $y$ dimension) through the FMI barriers at temperature $T$ is calculated from the Landauer-B\"{u}ttiker formalism,~\cite{Buttiker1985}
\begin{eqnarray}
G_{M_{j}}(E_{F},V)=\frac{e^{2}}{\pi \hbar }\int_{-\infty }^{\infty }dE\int_{-\frac{\pi }{2}}^{\frac{\pi }{2}}d\theta ~\mathrm{P_{T}}(E_{F},\theta,V
)F(E,E_{F},T)\,,
\label{eqn:Landauer}
\end{eqnarray}
where $F(E,E_{F},T)=-df(E,E_{F},T)/dE$ and $f(E,E_{F},T)$ is the Fermi-Dirac
distribution with Fermi energy $E_{F}$. $k_{x}$ and $k_{y}$ (thus, $k^B_x$ in region B) are substituted by $|E|\cos \theta $ and $ |E|\sin \theta $ with  incident angle $\theta $. The Dirac point is chosen as the reference (zero) for the energy coordinate (i.e., $E_F$, $E$) in the the present convention.

Based on this equation [Eq.~(\ref{eqn:Landauer})], the surface resistance $R_{M_{j}}$ is found by the reciprocal relation, $R_{M_{j}}~=~G_{M_{j}}^{-1}$, and $R_{M_{j}}$ is analyzed for three different magnetizations $\mathbf{M}$, each  directed along the axes $j=x$, $y$ and $z$.  Specifically, a 30-nm FMI strip is considered at room temperature as a function of $E_{F}$ and $V$. The exchange energy at the TI/FMI interface is assumed to be 40~meV, which appears practically attainable in the current technology.~\cite{Semenov2008}  The calculation results clearly indicate that $R_{M_{y}}$ ($j=y$) has the smallest resistance of the three for all values of $E_F$ and $V$ under investigation. Apparently the Dirac cone displaced in the $x$ direction (induced by $M_{y}$) does not significantly hinder electron transmission in the same direction in accord with the earlier qualitative analysis. On the other hand, $R_{M_{z}}$ ($j=z$) is the largest as expected (due to the bandgap, etc.).  As $E_F$ deviates from the Dirac point, the filtering effect of the FMI barrier appears to be reduced, leading to a gradual decrease in $R_{M_{x}}$ and $R_{M_{z}}$, while $R_{M_{y}}$ remains unaffected. This can be understood from the fact that the current contribution comes predominantly from electrons near the Fermi surface (i.e., where $df/dE \neq 0$).  Consequently, an increase of $E_F$ means a higher average energy for conducting electrons which are less affected by the "potential barriers". Thus, the spin filtering effect is the largest when $E_{F}$ is at the Dirac point. As for the dependence on the gate bias, it is found that $R_{M_{y}}$ increases with $V$ slightly. On the other hand, $R_{M_{z}}$ and $R_{M_{x}}$ decrease as $V$ becomes larger due to the Klein tunneling as discussed in Refs.~\onlinecite{Mondal2010} and \onlinecite{Wu2010}.  In fact, $R_{M_{y}}$ should also start to decrease at large $V$ (that is beyond the range of the current calculation).   The values of $R_{M_{z}}$, while they are close, stay generally higher than $R_{M_{x}}$ as indicated earlier. Since we consider the ideal case where the conduction and valence bands are symmetric, the resistance variations described above would also show a symmetric behavior for negative values of $E_F$ and $V$; they are expected to be asymmetric in real materials due to the nonlinearity.

Since the absolute values of the resistance are varied widely by specific parameters for materials and structures, the comparison between different magnetic configurations may be best achieved by their relative ratios as illustrated in Fig.~2. This is particularly the case when one of the quantities (i.e., $R_{M_y}$) is not very sensitive to both $E_F$ and $V$ in the range under consideration.  As the Fermi energy level increases from 0 to 50 meV, $R_{M_{z}}/R_{M_{y}}$ drops from approximately 9.1 to 3.2 while $R_{M_{x}}/R_{M_{y}}$ is reduced from 6.9 to 3.0. Similarly, $R_{M_{z}}/R_{M_{y}}$ decreases from 9.1 to 1.8 and $R_{M_{x}}/R_{M_{y}}$ from 6.9 to 1.3 with a gate bias sweep in the same range.  Adopting the definition of magnetoresistance $\mathrm{MR} =({R_{M_{i}}}/{R_{M_{j}}} -1)\times 100\%$, the results clearly suggest that $\mathrm{MR}$ as large as 800~\% may be achieved at room temperature in the proposed single FMI barrier-TI system with a 90$^{\circ }$ rotation of magnetization [e.g., $ \mathbf{M}$$\parallel$$y$ $ \leftrightarrow$ $\mathbf{M}$$\parallel$$z$, see Fig.~1(a)].

Note that the giant magnetoresistance is achieved with only one FMI
barrier.  This is possible because the electron spin and momentum are
locked in the manner of Bychkov-Rashba effect with an extremely strong
spin-orbital interaction $-$ one of the fascinating physical phenomena that a TI can provide. In conventional materials, it is necessary to set a reference magnetization for spin-based current manipulation. As such, the electric properties are varied by controlling the relative spin states between two or more magnetization directions.  Ferromagnetic metallic contacts are typically used for spin polarized carrier injection, and the current is controlled by another ferromagnetic metallic contact or gate.  In a TI system, on the contrary, the current direction itself can work as the reference for spin, eliminating the need for one of the magnetic layers.

%By the way, we found the minor resistance variation in a 180$^{\circ }$ %rotation of magnetization which is equivalent to $\mathbf{J}\rightarrow %-\mathbf{J}$.

For room temperature operation of the structure, the gating function should be
robust against the thermal broadening, which could wash out the effects of
the FMI barrier.  Accordingly, we examine the magnetoresistance as a function of exchange energy and temperature as summarized in Fig.~3.  Here, two sets of curves are plotted in terms of the exchange energy $A^{ex}$ for analysis.  One set represents MR at fixed temperatures (400, 300, 200, 100~K), while the other (thin dotted lines) consider it when both temperature and $A^{ex}$ vary but their ratio is kept constant (i.e., $A^{ex}/k_BT = n$). Generally speaking, MR increases with $A^{ex}$ while the temperature has a negative influence as expected. Interestingly, it appears that MR of 200~\% or larger is attainable by the $ \mathbf{M} $$ \parallel $$ y $-$ \mathbf{M} $$ \parallel $$ z $ rotation once $A^{ex}$ satisfies two conditions [Fig.~3(a)]; i.e., $A^{ex}$ is larger than the average electron thermal energy $k_{B}T$ and, at the same time, larger than the absolute value of 20~meV. The magnitude of MR increases to over 1000~\%, if $ A^{ex}$ is larger than twice the average thermal energy and larger than 20~meV. For instance, MR reaches approximately 1300~\% when $T=300$~K and $A^{ex} = 52$~meV. As for the $ \mathbf{M} $$ \parallel $$ y $-$ \mathbf{M} $$ \parallel $$ x $ rotation [Fig.~3(b)], the values of MR are generally lower than those of $ \mathbf{M} $$ \parallel $$ y $-$ \mathbf{M} $$ \parallel $$ z $ rotation in accord with the earlier discussion. However, MR of a few hundred percent can be obtained similarly when $A^{ex} > k_BT$.  This provides a clear indication that the predicted giant magnetoresistance effect could be sufficiently robust for practical use at room temperature.

The induced change in the channel resistance as a function of the
magnetization direction can find an immediate application in non-volatile
memory.  There are FMI materials with multiple metastable magnetization
states (i.e., directions) that can be used for information storage.  The surface states of a TI in contact would provide an efficient read-out medium.  Furthermore, one can envision a transistor-like function if dynamic control of magnetization direction is achieved as a part of the gate structure.  Such a concept may be realized by employing a multiferroic heterostructure;~\cite{Zavaliche2007,Lebeugle2009,Kim2004} that is, an articulated combination of ferromagnetic and piezo-/ferroelectric materials as the gate dielectric.  Then, the applied gate bias can modulate not only the electrostatic potential in the channel but also the surface state band structure itself (i.e., the $E-k$ relationship), leading potentially to a more drastic response in the drain current than that in conventional field-effect transistors (FETs).  The desired case is when the 90$^\circ$ rotation of magnetization ($ \mathbf{M} $$ \parallel $$ z $$\rightarrow$$ \mathbf{M} $$ \parallel $$ y $) is designed to coincide with the electrostatic turn-on of the device as schematically illustrated in Fig.~1(c).  Although the improvement in the on/off slope could be substantial in the scaled devices based on tunneling, the largest effect is expected for the intermediate channel lengths where the devices switch between the diffusive (low bias) and the ballistic (high bias) transport regimes.  Due to the break in the time reversal symmetry, it is very likely that the $ \mathbf{M} $$ \parallel $$ z $ case (with a parabolic band structure) possesses a much shorter electron mean free path than the other with a massless Dirac cone.  Hence, a judicious choice of the channel length could ensure a very significant change in the channel mobility in the on/off transition $-$ a highly sought-after feature for low-power devices.  Quantitative estimate of the potential advantage over the conventional FETs requires a detailed investigation of electron-phonon interactions~\cite{Boryse2011} in TIs and is beyond the scope of the current analysis.

%In summary, we studied the behavior of Dirac fermions in a TI structure with a %single FMI barrier. The results show that our TI structure has unique features %due to the spin polarized nature of the surface states, contrasting with %conventional materials. Particularly, it appears that giant magnetoresistance %can be achievable with proper manipulation of the magnetization direction of
%a single FMI strip. Thermalization effects on the transport properties were
%also analyzed for possible room temperature device applications. Based on
%our findings, we discussed a few possible devices and showed that dynamical
%control of conductance could be achievable by using a multiferroic
%heterostructure, which potentially works as voltage controlled switch with
%high mobility.

% If you have acknowledgments, this puts in the proper section head.

This work was supported, in part, by the SRC Focus Center on Functional
Engineered Nano Architectonics (FENA) and the US Army Research Office.

\clearpage \noindent Figure Captions

\vspace{0.5cm} \noindent Figure~1. (Color online)
Schematic illustration of the considered TI/FMI system. Rotation of the
magnetization direction around the x ($ \mathbf{M}$$\parallel$$y$ $ \leftrightarrow$ $\mathbf{M}$$\parallel$$z$) and z ($ \mathbf{M}$$\parallel$$y$ $ \leftrightarrow$ $\mathbf{M}$$\parallel$$x$) axes is shown in (a) and (b), respectively.  (c) Envisioned device operation of the proposed structure as an FET (drain current vs.\ gate bias).

\vspace{0.5cm} \noindent Figure~2. (Color online) Relative ratio of surface resistance as a function of (a) Fermi energy and (b) gate bias when the magnetization rotates 90$^\circ$ ($ \mathbf{M}$$\parallel$$y$ $ \leftrightarrow$ $\mathbf{M}$$\parallel$$\{x,z\}$).  The solid lines represent the values of $R_{M_z} / R_{M_y}$ [Fig.~1(a)], while the dashed lines show those of $ R_{M_x} / R_{M_y}$ [Fig.~1(b)].  A 30-nm FMI strip is considered at room temperature along with the exchange energy of 40~meV at the TI/FMI interface.

\vspace{0.5cm} \noindent Figure~3. (Color online)
Magnetoresistance $\mathrm{MR}$ [$ =({R_{M_{i}}}/{R_{M_{j}}} -1)\times 100\%$] as a function of exchange energy $A^{ex}$ and temperature $T$. (a) and (b) correspond to ($i$=$z$,$j$=$y$) and ($i$=$x$,$j$=$y$), respectively. For both (a) and (b), thick dashed, solid, dotted, and dash-dotted lines represent the results at constant temperature $T=$~400, 300, 200, 100~K. On the other hand, thin dotted curves are obtained with the $A^{ex}$ to $T$ ratio kept constant (i.e., $A^{ex}/k_BT = n$, where $n=1$, $2$, and $3$ as denoted in the figure).

\clearpage

\begin{center}
\begin{figure}[tbp]
\includegraphics [bb=4 14 226 163,width=8cm]{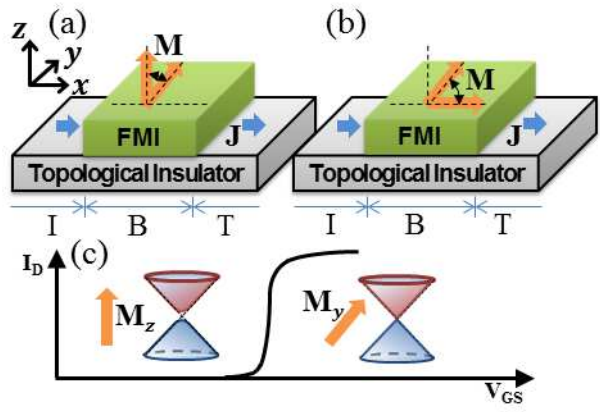} %For Arxiv
\caption{Kong et al.}
%\label{fig:schematic}
\end{figure}
\end{center}

\clearpage

\begin{center}
\begin{figure}[tbp]
\includegraphics[bb=3 2 221 154,width=8cm]{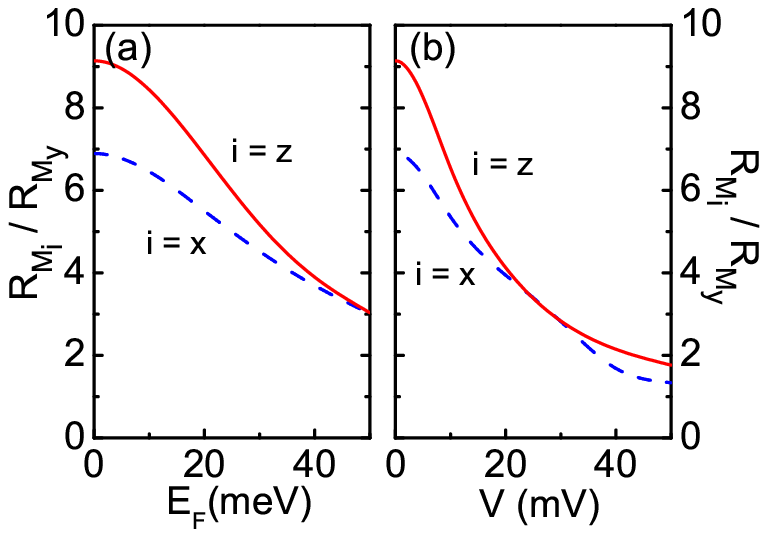}
\caption{Kong et al.}
%\label{fig:ratio}
\end{figure}
\end{center}

\clearpage

\begin{center}
\begin{figure}[tbp]
\includegraphics[bb=4 9 224 153,width=8cm]{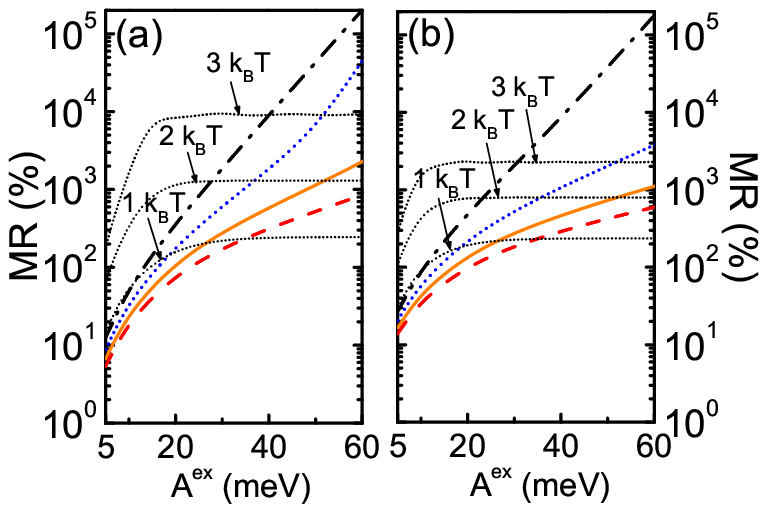}
\caption{Kong et al.}
%\label{fig:mr}
\end{figure}
\end{center}

\end{document}